\documentclass[twocolumn]{aastex631}

\newcommand{\Msun}{\,{\rm M_\odot}}
\newcommand{\fedd}{\,{f_{\rm Edd}}}
\newcommand{\Mblack}{M_\bullet}

\usepackage{booktabs}
\usepackage{amsmath}

\begin{document}

\title{Constraining Wind-Driven Accretion Onto Gaia BH3 With Chandra}

\author[0000-0002-1697-186X]{Nico Cappelluti}
\affiliation{Department of Physics, University of Miami, 
Coral Gables, FL 33124, USA}

\author[0000-0001-9879-7780]{Fabio Pacucci}
\affiliation{Center for Astrophysics $\vert$ Harvard \& Smithsonian, Cambridge, MA 02138, USA}
\affiliation{Black Hole Initiative, Harvard University, Cambridge, MA 02138, USA}

\author[0000-0002-0797-0646]{G\"unther Hasinger}
\affiliation{TU Dresden, Institute of Nuclear and Particle Physics, 01062 Dresden, Germany}
\affiliation{DESY, Notkestra{\ss}e 85, 22607 Hamburg, Germany}
\affiliation{Deutsches Zentrum für Astrophysik, Postplatz 1, 02826 Görlitz, Germany}



\begin{abstract}
Gaia BH3 is the most massive known stellar-origin black hole in the Milky Way, with a mass $\Mblack \approx 33 \Msun$. Detected from Gaia's astrometry, this black hole is in the mass range of those observed via gravitational waves, whose nature is still highly debated. Hosted in a binary system with a companion giant star that is too far away for Roche-lobe mass transfer, this black hole could nonetheless accrete at low levels due to wind-driven mass-loss from its companion star, thus accreting in advection-dominated accretion flow, or ADAF, mode. Using stellar wind models, we constrain its Eddington ratio in the range $10^{-9} < \fedd < 10^{-7}$, corresponding to radiative efficiencies $5\times10^{-5} < \epsilon < 10^{-3}$, compatible with radiatively inefficient accretion modes. Chandra ACIS-S observed this object and obtained the most sensitive upper bound of its [2-10] keV flux: $F_X < 3.25 \times 10^{-15} \, \rm erg \, s^{-1} \, cm^{-2}$ at $90\%$ confidence level corresponding to $L_{[2-10]}< 2.10 \times 10^{29} \, \rm erg \, s^{-1}$. Using ADAF emission models, we constrained its accretion rate to $\fedd< 4.91 \times 10^{-7}$ at the apastron, in agreement with our theoretical estimate. At the periastron, we expect fluxes $\sim 50$ times larger. Because of the inferred low rates, accretion did not significantly contribute to black hole growth over the system's lifetime. 
Detecting the electromagnetic emission from Gaia BH3 will be fundamental to informing stellar wind and accretion disk models.
\end{abstract}



\section{Introduction} \label{sec:intro}
Astrometry from the Gaia mission \citep{GAIA_2016} recently led to the detection of Gaia BH3 \citep{GBH3}. With a dynamically estimated mass of $\Mblack \approx 33 \Msun$, this black hole (BH) is the most massive known with a clear stellar origin.
Considering its relatively short distance of $590.6 \pm 5.8$ pc, Gaia BH3 represents a unique opportunity to study BHs of stellar origin. The mass of this BH is of particular importance, as it is similar to that of systems discovered by LIGO, which have merger masses in the range of $20-100 \Msun$. Also Cyg X-1, with a mass of $\approx 21 \Msun$, belongs to this range \citep{Miller-Jones_2021}. Remarkably, Gaia BH3 is characterized by a mass similar to that of the progenitors of the first gravitational wave detected by LIGO, which ended up forming a $62 \Msun$ BH \citep{LIGO_2016}. Thus, Gaia BH3 bridges the gap between gravitational wave sources and more traditional electromagnetic observations.

Considering its relatively short distance of 590.6 +/- 5.8 pc, Gaia BH3 represents a unique opportunity to study BHs of stellar origin. 

The detection of Gaia BH3 follows that of Gaia BH1 \citep{Gaia_BH1} and Gaia BH2 \citep{Gaia_BH2}. The first two systems discovered were remarkably similar, each hosting a $\sim 9 \Msun$ BH. Hence, Gaia BH3 represents a leap of a factor $\gtrsim 3$ upward in mass.

Gaia Data Release 3 \citep[DR3,][]{GDR2} preliminary data, supplemented by spectroscopy, has revealed for Gaia BH3 a binary system comprising a dark object and a very metal-poor giant companion with a mass of $0.76 \pm 0.05 \Msun$ with an orbital period of $11.6$ yr and a semi-major axis of $\sim 16$ AU. The companion star is caught in a late evolutionary period, ascending the giant branch \citep{GBH3}.

Stellar-mass BHs can be detected electromagnetically in close binaries, where the companion star is sufficiently close to build up a significant accretion rate onto the compact object \citep{McClintock_2006}.
Mass transfer can occur via Roche lobe overflow (the classic low-mass X-ray binary picture, or LMXB) or wind accretion (the classic high-mass X-ray binary picture, or HMXB). Classic HMXB systems are Cygnus X-1 (which can be at a permanent ``high'' state) and V4641 Sgr. Typically, LMXBs can undergo sudden outbursts in X-ray luminosity, named X-ray novae \citep{Remillard_2006}.

At a distance of $\sim 27$ AU (i.e., the companion star is currently at the apoapsis from the BH), the star is too far for mass transfer via the first Lagrangian point. Nonetheless, if sufficiently strong wind-driven outflows characterize the companion star, and if a non-negligible fraction of them falls onto the BH, then Gaia BH3 may be bright enough to become electromagnetically detectable by current or future observatories, as was previously argued for Gaia BH1 and BH2 \citep{Non_detections_2023}.

Previously, SRG/eROSITA scanned the sky region containing Gaia BH3 on four occasions during its all-sky survey in 2020-2022. No X-ray source was detected at its position, setting an upper limit on the [0.3-2.3]([2-10]) keV flux of $1.2 (2.6) \times 10^{-14} \, \rm erg \, s^{-1} \, cm^{-2}$ \citep{gilf}. At a distance of $590$ pc, this corresponds to an upper limit on the intrinsics X-ray luminosity of $5.0(10.8) \times 10^{29} \, \rm erg \, s^{-1}$. Note that the (bolometric) Eddington luminosity for a $33 \Msun$ BH is $4.2\times 10^{39}  \, \rm erg \, s^{-1}$. Hence, depending on the specific X-ray bolometric correction, this BH should be shining at a luminosity level $\ll 10^{-10} \times L_{\rm Edd}$, consistent with very weak accretion.

The eROSITA luminosity upper limit is consistent with weak accretion activity onto Gaia BH3. 
Furthermore, Chandra previously observed Gaia BH1 \citep{Gaia_BH1} and Gaia BH2 \citep{Gaia_BH2}, resulting in X-ray (and radio) non-detections \citep{Non_detections_2023}. 
In the case of Gaia BH3, the fact that the compact component is $\gtrsim 3$ times more massive and the companion star is evolved (i.e., with more significant winds than main-sequence stars) may fuel sufficient accretion via the Bondi-Hoyle-Lyttleton mechanism \citep{Bondi_1952} to make it electromagnetically detectable.

Based on the significant separation from the companion star and the upper limits already established by eROSITA, we expect Gaia BH3 to be accreting in advection-dominated accretion flow mode, or ADAF \citep{Narayan_1994, Narayan_1995, Abramowicz_1995, Narayan_2008, Yuan_Narayan_2014}. For Eddington ratios $\fedd \ll 10^{-2}$ (where the Eddington ratio is defined as the actual accretion rate normalized to the Eddington rate), the accretion disk onto compact objects becomes very radiatively inefficient, thus departing from the standard value of $\epsilon \sim 0.1$, which regulates the transformation of matter into radiation: $L = \epsilon \dot{M} c^2$. Hence, constraining the accretion properties of Gaia BH3 will require careful modeling of compact objects accreting in radiatively inefficient modes.

Determining or constraining Gaia BH3's current accretion properties will enable the reconstruction of its growth rate and, hence, its mass growth history. In other words, was its mass mainly acquired at formation or via mass growth? Considering the orbital and stellar properties of the system, it is likely that no significant modifications to the status quo have occurred in the last several billion years.

Gaia BH3's mass exceeds $20 \Msun$, a threshold not previously detected in Gaia data or high-mass X-ray binaries (HMXB, \citealt{McClintock_2006}). 
Hence, this discovery supports the scenario where high-mass BHs are remnants of low-metallicity stars \citep[as suggested by][]{bel16}. The low metallicity of Gaia BH3's companion star (i.e., [M/H]$<$-2, \citealt{GBH3}) suggests that high-mass BHs form predominantly at very low metallicities.

Generally speaking, the formation scenarios for Gaia BH3's system are complex and likely involve more than isolated binary evolution. For instance, \cite{elb} argued that Gaia BH3 could not have formed via isolated binary evolution channels, possibly requiring dynamical interactions in a dense environment. This hypothesis is potentially supported by Gaia BH3's association with the ED-2 stream, a possible remnant of a globular cluster \citep{Dodd_2023, Balbinot_2023, balb24}. On the other hand, \citealt{iorio} supports the thesis of an isolated binary history. One of the variables at play in discerning the nature of this binary system and of the BH itself is the native kick that can influence the system's parameters. Additionally, Gaia BH3 could be a primordial BH \cite[see, e.g.][for a review]{carr24} that dynamically captured the companion star in the ED-2 cluster. While this last hypothesis has not been investigated yet, Gaia BH3 falls in the mass range where PBH could contribute to a non-negligible fraction of dark matter \citep[see, e.g.][]{gb17}

This paper aims to further explore the nature of Gaia BH3 via X-ray and UV observations with Chandra and Swift-UVOT. In particular. We present the results of a $10$ ks Director Discretionary Time (DDT) and a 1.3 ks Swift ToO observations associated with carefully modeling its accretion properties via ADAF models.
\vspace{1cm}

\section{Data Analysis}
\begin{figure*}[t]
\includegraphics[width=0.98\textwidth]{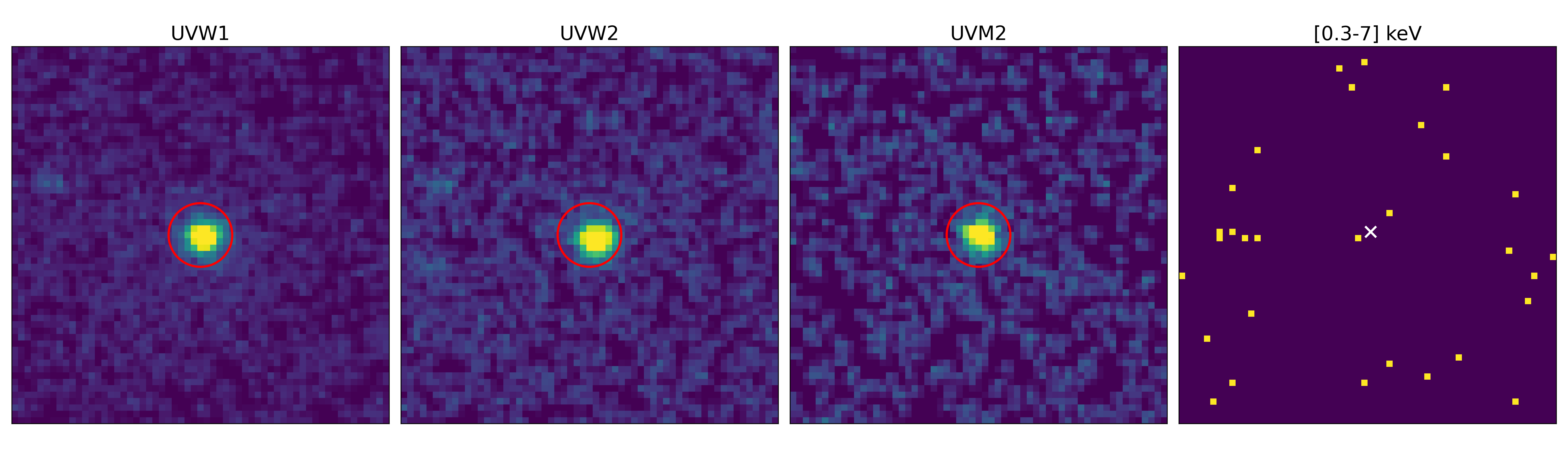}
\caption{\label{fig:image} From left to right: Swift-UVOT UVW1, UVW2, UVM2 and Chandra [0.3-7] keV image, respectively (60$^{\prime\prime}\times$60$^{\prime\prime}$) centered on the position of Gaia BH3. The pixel scale is 1$^{\prime\prime}$/pix. The red circle indicates a radius of 5$^{\prime\prime}$ around the best astrometric solution from Gaia. In the X-ray image we show with a white $X$ the position of the Gaia-BH3 system. The size of the symbol represents the Chandra astrometric accuracy.}
\end{figure*}

Gaia BH3 (RA: 19:39:18.69, Dec: +14:55:51.53) was observed with Chandra ACIS-S on May 25, 2024. The observation (ID: 29396, PI: Pacucci) was taken in VFAINT telemetry mode and TE Exposure Mode. 
Level 2 grade event files were created and calibrated using the Chandra CIAO tool chandra\_repro and removed the time interval affected by background flares, thus obtaining a final net exposure of 9.95 ks. No astrometric correction was applied since no bright sources were available in the FOV. The source was placed on the optical axis, hence observed at the nominal angular resolution of $\sim$ 0.5$^{\prime\prime}$, which is of the same order of magnitude as the absolute astrometric precision of the instrument. 
We created a [0.3-7] keV (full) band image and searched within a radius of 2$^{\prime\prime}$ around the best astrometric solution for the Gaia BH3 system from Gaia. A simple eye inspection revealed no counts within the search region. The image of the on-axis region of our observation is shown in Fig. \ref{fig:image}.

We then calculated the upper bounds for the source flux by using the method developed by \cite{primini} and streamlined in the CIAO package {\texttt srcflux}. The method first extracts source photons within 90\% of the PSF-encircled energy fraction and background photons in an annulus around the source. Counts are then converted into count rates by applying an exposure map. Then, using a Bayesian approach, 90\% confidence limits on the source flux are obtained after converting count rates by assuming a spectrum modeled with a galactic obscured \citep[i.e., Log $N_H = 21.5 \rm \, cm^{-2}$,][]{2005A&A...440..775K} power-law with a spectral index $\gamma$=2. 

Since no counts were detected in the search area, we find a $90\%$ confidence upper bound for the [0.5-7] keV count rate of 2.67$\times$10$^{-3} \, \rm cts \, s^{-1}$. The $10$ ks exposure and the local background rate correspond to an upper limit of $2.67$ net counts, which translates into a [0.5-7]([2-10]) absorbed flux of $3.76(3.25) \times 10^{-15} \, \rm erg \, s^{-1} \, cm^{-2}$. Correcting for the absorption, the flux is $5.86(5.06) \times 10^{-15} \, \rm erg \, s^{-1} \, cm^{-2}$, which translates to a [0.5-7]([2-10]) keV intrinsic luminosity L$_X$=$2.44(2.10) \times 10^{29} \, \rm erg \, s^{-1}$. This value is $\sim 10$ times below the estimated [2-10] keV upper limit by eROSITA. These values are summarized in Table \ref{tab:xray_limits}.

{\em Swift} observed Gaia-BH3 on 04-19-2024 (OBS-ID 00016607001) for a total of 1.3 ks split over the U,B,V, UVW1, UVW2 and UVM2 filters. The source was saturated in the U, B, and V bands; therefore, we focus on the previously unavailable UV data collected in the UVW1, UVW2, and UVM2 filters. 
Cleaned data from {\em Swift} ToO observations are made public immediately; therefore, we simply performed aperture photometry using the Swift tool {\em uvotsource}. We imposed a 5$^{\prime\prime}$ extraction radius around the coordinate of GAIA BH3 and selected an adjacent empty region of twice the radius to evaluate the background. After extraction, data are calibrated using the most recent CALDB, provided by HEASOFT. In all three bands, the source (the companion star) is detected at high significance (i.e., $>15\sigma$), and the corresponding fluxes are reported in Tab. \ref{tab:xray_limits}.

In order to properly evaluate the contamination from the companion star, we queried the Vizie-r \citep{vizier} photometry service within 0.3$^{\prime\prime}$ from the Gaia coordinates. We tried larger radii but the results were severely contaminated by neighbouring stars. We collected data from the mid-IR to the optical band from a combination of ground and space-based catalogs \citep[see,][data reported in Fig. \ref{fig:SED}]{2004AJ....127.3043Z,2010AJ....139.2440R,2017A&A...600L...4A,2017AJ....153..166Z,2019AJ....158..138S,2021arXiv210804778P,2018AJ....156..102S,2016MNRAS.463.4210N}. 
Fig. 2 shows that the UV flux of the system can be entirely attributed to the companion star of GAIA-BH3 (see below for a thorough discussion of the BH SED).

\begin{table*}
\centering
\caption{90\% confidence upper limits on X-ray Emission from Gaia BH3.}
\label{tab:xray_limits}
\begin{tabular}{@{}ccccc@{}}
\toprule
  & & {\em Chandra ACIS-S} & & \\
\midrule
Energy Range & Observed F$_X$   &  Unobscured F$_X$ & Intrisic L$_X$ & Periastron F$_X$  \\ 
  keV   &  erg cm$^{-2}$ s$^{-1}$   & erg cm$^{-2}$ s$^{-1}$ &  erg s$^{-1}$& erg cm$^{-2}$ s$^{-1}$ \\
\midrule
0.5-7             & $<3.76 \times 10^{-15}$        &      $<5.86 \times 10^{-15}$       & $<2.44 \times 10^{29}$    &      $<1.52 \times 10^{-13}$           \\
2-10              & $<3.25 \times 10^{-15}$         &     $<5.06 \times 10^{-15}$         & $<2.10 \times 10^{29}$      & $<1.31 \times 10^{-13}$           \\ \midrule
 & & {\em Swift UVOT} & & \\
\midrule

Frequency & Flux   & m$_{AB}$ &FILTER &    \\ 
  Hz   &  $mJy$   &    &  &  \\
   \midrule
1.157$\times$10$^{15}$&2.90$\pm{0.09}$ &  15.23$\pm{0.03}$ &UVW1  & \\
1.345$\times$10$^{15}$&0.54$\pm{0.03}$ &16.88$\pm{0.04}$ & UVW2  & \\
1.457$\times$10$^{15}$&0.64$\pm{0.03}$ & 17.07$\pm{0.07}$  & UVM2  & \\
\bottomrule
\end{tabular}
\end{table*}

\section{Modeling the Accretion by Wind-Driven Mass Loss from the Companion Star}

Two accretion avenues are available to feed Gaia BH3 with enough gas to render it electromagnetically detectable: (i) from the ISM and (ii) from wind-driven mass loss from the nearby companion star.

BHs wandering in the Milky Way Galaxy may cross ISM environments dense enough to produce a non-negligible accretion rate \citep{Agol_2002}. This avenue was recently explored in depth by several studies (see, e.g., \citealt{Seepaul_2022}), also for the case of Gaia BH1 and BH2 \citep{Non_detections_2023}; detailed predictions for future X-ray observatories were also made \citep{AXIS_WANDERING}. The accretion level reachable from the ISM is strongly dependent on the density of the environment that the BH is crossing and the relative velocity with respect to it, as expressed by the standard Bondi-Hoyle-Lyttleton accretion model \citep{Lyttleton_1939, Bondi_1944, Bondi_1952}:
\begin{equation}
    \dot{M}_{\rm B} = \frac{4 \pi G^{2} \Mblack^{2} \rho}{(v_{\rm rel}^{2} + c_{s}^{2})^{3/2}} \, .
    \label{eq:Bondi_basic}
\end{equation}
In this equation, $\rho$ and $c_{s}$ are the density and sound speed of the ambient ISM, and $v_{\rm rel}$ is the relative speed with the ISM.

Assuming the passage within a dense molecular cloud, with a number density of $10^3 \, \rm cm^{-3}$, and a favorable (i.e., low) relative velocity, X-ray fluxes up to $10^{-6} \, \rm erg \, s^{-1} \, cm^{-2}$ could occur. However, dense molecular clouds occupy only $\sim 0.05\%$ of volume in the Milky Way \citep{Ferriere_2001}; hence, such an encounter is unlikely. A passage within the most likely environment, the hot ionized medium, would lead to fainter X-ray fluxes by $10-15$ orders of magnitude, i.e., $10^{-16}-10^{-21} \, \rm erg \, s^{-1} \, cm^{-2}$.
Given that this feeding mechanism strongly depends on the environment, we focus on the second channel, which depends on an element that is certainly there: the companion star.

\subsection{Theoretical Estimates}
\label{subsec:theory_estimates}
In this Section, we estimate (i) the wind-driven mass-loss rate from the companion star, (ii) the wind velocity, (iii) the Bondi-Hoyle-Lyttleton accretion rate on the BH, and (iv) the radiative efficiency of the accretion process. In the next Section, we compare these estimates with constraints from our Chandra observation.

Gaia BH3 orbits an old, very metal-poor, giant star, ascending the red giant branch \citep{GBH3}. Giant stars have significantly larger wind-driven outflow rates than the solar value (i.e., $\sim 2 \times 10^{-14} \rm \Msun \, yr^{-1}$, see, e.g., \citealt{Johnstone_2015}). 
However, significant uncertainties exist in determining mass-loss rates, except for the Sun, for which we can directly measure the value of its wind. Even for the Sun, its wind activity is subject to time variations depending on the solar cycle and other factors (see, e.g., \citealt{Wang_1998}). 

Several studies have assembled values of measured mass-loss rates from the literature and created best-fit relations as a function of relevant stellar parameters. For example, the classic relation introduced by \cite{Reimers_1975} for red giant stars is the following:
\begin{equation}
  \dot{M}_{\text {w}}=4 \times 10^{-13} \beta_R\left(\frac{L_{\star}}{L_{\odot}}\right)\left(\frac{R_{\star}}{R_{\odot}}\right)\left(\frac{M_{\star}}{M_{\odot}}\right)^{-1} \Msun \, \mathrm{yr}^{-1} \, ,
\end{equation}
where $L_{\star}$, $R_{\star}$, and $M_{\star}$ are stellar parameters, and $\beta_R$ is a scaling factor which, for red giant branch stars, is $\beta_R \sim 0.1$ \citep{Reimers_1975, Non_detections_2023}. 
A more recent mass-loss relation, introduced by \cite{Meszaros_2009}, takes into account the metallicity of the star:
\begin{equation}
    \dot{M}_{\text {w}}=0.092 \times \mathrm{L}^{0.16} \times \mathrm{T}_{\mathrm{eff}}^{-2.02} \times \mathrm{A}^{0.37} \Msun \, \mathrm{yr}^{-1} \, ,
\end{equation}
where $\mathrm{A}=10^{[\rm{Fe/H}]}$.
Using the stellar values appropriate for Gaia BH3's companion, we find mass-loss rates of $\dot{M}_{\text {w}} \sim 10^{-11} \, \Msun \, \mathrm{yr}^{-1}$ using \cite{Reimers_1975} and $\dot{M}_{\text {w}} \sim 10^{-9} \, \Msun \, \mathrm{yr}^{-1}$ using \cite{Meszaros_2009}. Note that typical ranges for metal-poor field giants are between $3\times 10^{-10} \, \rm \Msun \, yr^{-1}$ and $6\times 10^{-8} \, \rm \Msun \, yr^{-1}$ \citep{Dupree_2009, Mullan_2019}.

The same considerations for mass-loss rates apply to the wind velocity, which is also challenging to estimate. Using the relation based on stellar parameters also employed in \cite{Non_detections_2023}:
\begin{equation}
    v_{\text {wind }}=600 \beta_{\text {wind }}\left(\frac{M_{\star}}{M_{\odot}}\right)^{1 / 2}\left(\frac{R_{\star}}{R_{\odot}}\right)^{-1 / 2} \mathrm{~km} \mathrm{~s}^{-1} \, ,
\end{equation}
where $\beta_{\text {wind}}$ is a scaling parameter. Here, we set $\beta_{\text {wind}} = 1$ and we derive a wind velocity of $v_{\text {wind}} \sim 200 \mathrm{~km} \mathrm{~s}^{-1}$, which is consistent with magneto-hydrodynamical simulations of wind velocities in the early phases of red giant branch evolution \citep{Yasuda_2019}.

We can estimate the Bondi-Hoyle-Lyttleton accretion rate once the wind-driven mass-loss rate and the wind velocities are estimated. In Eq. \ref{eq:Bondi_basic}, by assuming that the relative velocity $v$ (which, in practice, is the wind speed) is much higher than the sound speed of the medium $c_s$, we can derive a more appropriate form for the Bondi-Hoyle-Lyttleton accretion rate:
\begin{equation}
\dot{M}_{\mathrm{B}}=\frac{G^2 \Mblack^2 \dot{M}_{\text {wind }}}{v_{\text {wind }}^4 d_{\star \bullet}^2} \, ,
\label{eq:Bondi_new}
\end{equation}
where $d_{\star \bullet}$ is the distance between the BH and its companion star.
Using the two estimates of the mass-loss rates described above, we obtain the following range of accretion rates:
\begin{equation}
    7 \times 10^{-15} < \dot{M}_{\mathrm{B}} [\rm \Msun \, yr^{-1}] < 7 \times 10^{-13} \, .
\end{equation}
It is interesting to note that \textit{any value of mass accretion rate within this range would have contributed a mass $\Delta \Mblack \ll 1 \Msun$ to Gaia BH3 during the entire history of the Universe}.

Considering the Eddington accretion rate for a $33 \Msun$ BH, $\dot{M}_{\rm Edd} = 7.3 \times 10^{-6} \, \rm \Msun \, yr^{-1}$, we can estimate a range of Eddington ratios $\fedd$:
\begin{equation}
    10^{-9} < \fedd < 10^{-7}
    \label{eq:fedd_estimate}
\end{equation}

As mentioned in Sec. \ref{sec:intro}, for Eddington ratios $\fedd \ll 10^{-2}$, the compact object enters the accretion regime of ADAF \citep{Narayan_1994, Narayan_1995, Abramowicz_1995, Narayan_2008, Yuan_Narayan_2014}, with matter-to-energy efficiencies $\epsilon$ much lower than the radiative efficient value of $\sim 10\%$.
In particular, for hot accretion flows such as ADAF, we use the model presented in \cite{Xie_2012} to estimate the value of the matter-to-energy conversion factor $\epsilon$, defined as $\epsilon = L/(\dot{M} c^2)$:
\begin{equation}
    \epsilon = \epsilon_0 \left( 100 \times \fedd \right)^a \, ,
\end{equation}
where, for $\fedd \lesssim 10^{-5}$, we use $\epsilon_0 = 1.58$ and $a = 0.65$.
Hence, we obtain a range of radiative efficiencies:
\begin{equation}
    5\times 10^{-5} < \epsilon < 10^{-3} \, ,
\end{equation}
which are, in fact, much lower than the standard value for radiatively efficient flows.

The effective accretion rate falling onto the BH may be even lower than estimated. BHs accreting in ADAF mode are well-known to be characterized by strong winds, allowing typically $\lesssim 1\%$ of the gas
available at the Bondi radius to be actually accreted \citep{Yuan_2012, Yuan_2014}.
Over time, numerous adjustments to the standard Bondi model have been incorporated to account for effects that effectively limit the accretion rate, such as the presence of turbulent outflows \citep{Blandford_Begelman_1999} and convection within the accretion disk \citep{Narayan_2000, Quataert_Gruzinov_2000}.
To incorporate these factors, the standard Bondi accretion rate is modified as follows \citep{Igumenshchev_2003, Proga_2003}:
\begin{equation}
\dot{M} = \left(\frac{R_{\rm in}}{R_{\rm B}}\right)^{p} \dot{M}_{\rm B} \, ,
\label{eq:Bondi_corr}
\end{equation}
where we adopt $R_{\rm in} = 50 \times R_{\rm S}$ \citep{Abramowicz_2002} as the effective inner radius, where $R_{\rm S}$ is the Schwarzschild radius, and $R_{\rm B}$ represents the accretion radius or Bondi radius.
The parameter $p$ determines the transition of the accretion flow to sub-Bondi rates at radii smaller than the Bondi radius, and we set $p=0.5$, in line with simulations \citep{Pen_2003, Yuan_2012, Yuan_Narayan_2014, Ressler_2020}.
Notably, at radii $r > R_{\rm B}$, the accretion rate approximates the Bondi rate, whereas at $r < R_{\rm B}$, the accretion rate diminishes due to turbulent outflows and convection. Using this formalism, we derive that effective accretion rates onto the BH could be a factor $\sim 100$ lower.

A direct application of Eq. \ref{eq:Bondi_basic} for calculating the accretion rate in the case of Gaia BH2, combined with using wind models from \citealt{Meszaros_2009} would have made that BH possibly detectable in the X-rays (note that \citealt{Non_detections_2023} used the wind model from \citealt{Reimers_1975}). The lack of X-ray detection \citep{Non_detections_2023} perhaps signifies that strong ADAF winds are crucial in these systems. For this reason, in Sec. \ref{sec:future} we parametrize the accretion rate in terms of two general parameters only: the mass loss rate from the companion star and the fraction of available gas that is accreted onto the BH.

However, we also note that in Eq. \ref{eq:Bondi_basic}, often the velocity in the denominator is estimated as $v = \sqrt{v_{\rm rel} c_s} = \sqrt{v_{\rm wind} c_s}$. Assuming a typical value of $c_s = 10 \, \rm km \, s^{-1}$, we obtain a relative velocity $v_{\rm rel}$ that is $\sim 5$ times lower than the assumed wind speed, which would in turn lead to a significantly higher value of the accretion rate, given the steep dependence from $v_{\rm rel}$ in Eq. \ref{eq:Bondi_basic}. In summary, significant uncertainties affect our estimates of the Eddington ratio, strengthening the case for attempting a detection of this source.  

\subsection{Estimates Based on the Chandra Non-Detection}

Based on the unobscured X-ray flux upper limit of Gaia BH3 determined by our Chandra observation (see Table \ref{tab:xray_limits}), we estimate its full SED from radio to gamma rays and obtain an observational estimate of the Eddington ratio $\fedd$.

We use the analytical model for ADAF SEDs developed in \cite{Pesce_2021}, based on the original formalism by \cite{Mahadevan_1997}.
The model used to derive the analytical SEDs is based on self-similar advection-dominated accretion flows, which describe the local properties of the accreting gas as functions of mass, accretion rate, radius, viscosity parameter, and other factors. The accreting gas in this regime is a two-temperature, optically thin plasma where ions are at their virial temperature and electrons are cooler. The pressure is a combination of gas and magnetic pressures. The three emission mechanisms are synchrotron, inverse Compton, and bremsstrahlung \citep{Mahadevan_1997}.

This analytical model requires as inputs the BH mass ($\Mblack = 33 \Msun$) and the Eddington ratio $\fedd$ and provides an estimate of the SED based \citep{Pesce_2021}.
While the model is scale-free and can be used with any BH mass, it requires $\fedd < 10^{-1.7}$ (i.e., in the ADAF regime), which is well-suited for our purposes.

\begin{figure}[h]
    \centering
    \includegraphics[width=0.45\textwidth]{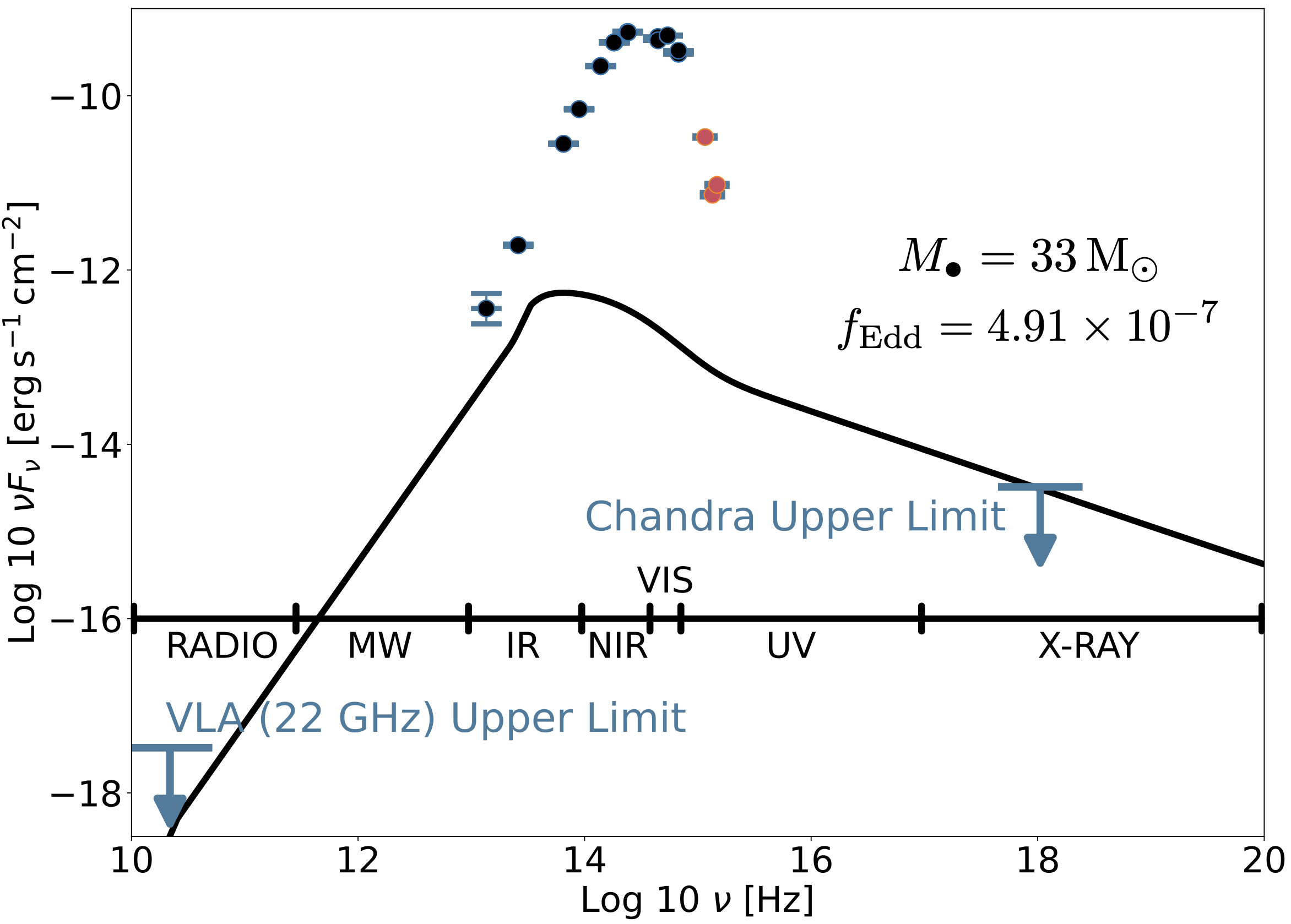}
    \caption{The solid black line shows the SED for an ADAF-accreting BH of $33 \Msun$, generated with the model described in \cite{Mahadevan_1997, Pesce_2021}. The X-ray fluxes in the SED are set to equal the upper limit determined from our Chandra observation. Hence, this model predicts that Gaia BH3 accretes with an Eddington ratio $\fedd < 4.91\times 10^{-7}$. The data points shown with their $1\sigma$ error bar demonstrate that the total emission of the Gaia BH3 system is dominated by the red giant; the points shown in red are from Swift UVOT. The VLA upper limit at 22 GHz is also shown \citep{VLA_upper_2024}.}
    \label{fig:SED}
\end{figure}

Given that the BH mass of Gaia BH3 is well constrained, we are left with only one free parameter: $\fedd$. We vary $\fedd$ until we obtain an X-ray flux equal to our Chandra-determined upper limits.
This result is obtained for $\fedd = 4.91 \times 10^{-7}$, and the resulting SED is shown in Fig. \ref{fig:SED}. We note that this SED is useful in estimating the emission of Gaia BH3 in other bands (e.g., radio or infrared, where it is expected to peak). 

Hence, we conclude that our Chandra observation provides the following upper limit:
\begin{equation}
    \fedd < 4.91\times 10^{-7} \, ,
\end{equation}
which is perfectly compatible with the previously estimated range $10^{-9} < \fedd < 10^{-7}$ (see Eq. \ref{eq:fedd_estimate}).

\section{Orbital Variations of the Accretion Rate: a  Look Into the Future} 
\label{sec:future}

In this Section, we provide a broader perspective on the possibility of observing Gaia BH3 in the future.

We start by commenting on the possibility of detecting Gaia BH3 at a given $\fedd$, with current and future X-ray observatories.
Since the BH mass and the distance to Gaia BH3 are fixed (at least at a level relevant for this work), the possibility of detecting it in the X-rays depends fundamentally on its accretion rate, i.e., to its Eddington ratio.
The Eddington ratio depends on two parameters: (i) the mass-loss rate of the companion star and (ii) the fraction of mass that actually falls onto the BH. The second parameter depends, in turn, on various factors such as the wind velocity, the reduction in accretion rate due to ADAF-related effects, etc.
For simplicity, we fold all these unknowns into the fraction ${\cal F}$, which represents the fraction of the mass-loss rate ${\cal \dot{M}}$ that falls onto the BH.
Hence, the Eddington ratio $\fedd$ is calculated as:
\begin{equation}
    \fedd = \frac{ {\cal \dot{M}} {\cal F} }{\dot{M}_{\rm Edd}} \, .
\end{equation}

\begin{figure}
    \centering
    \includegraphics[width=0.49\textwidth]{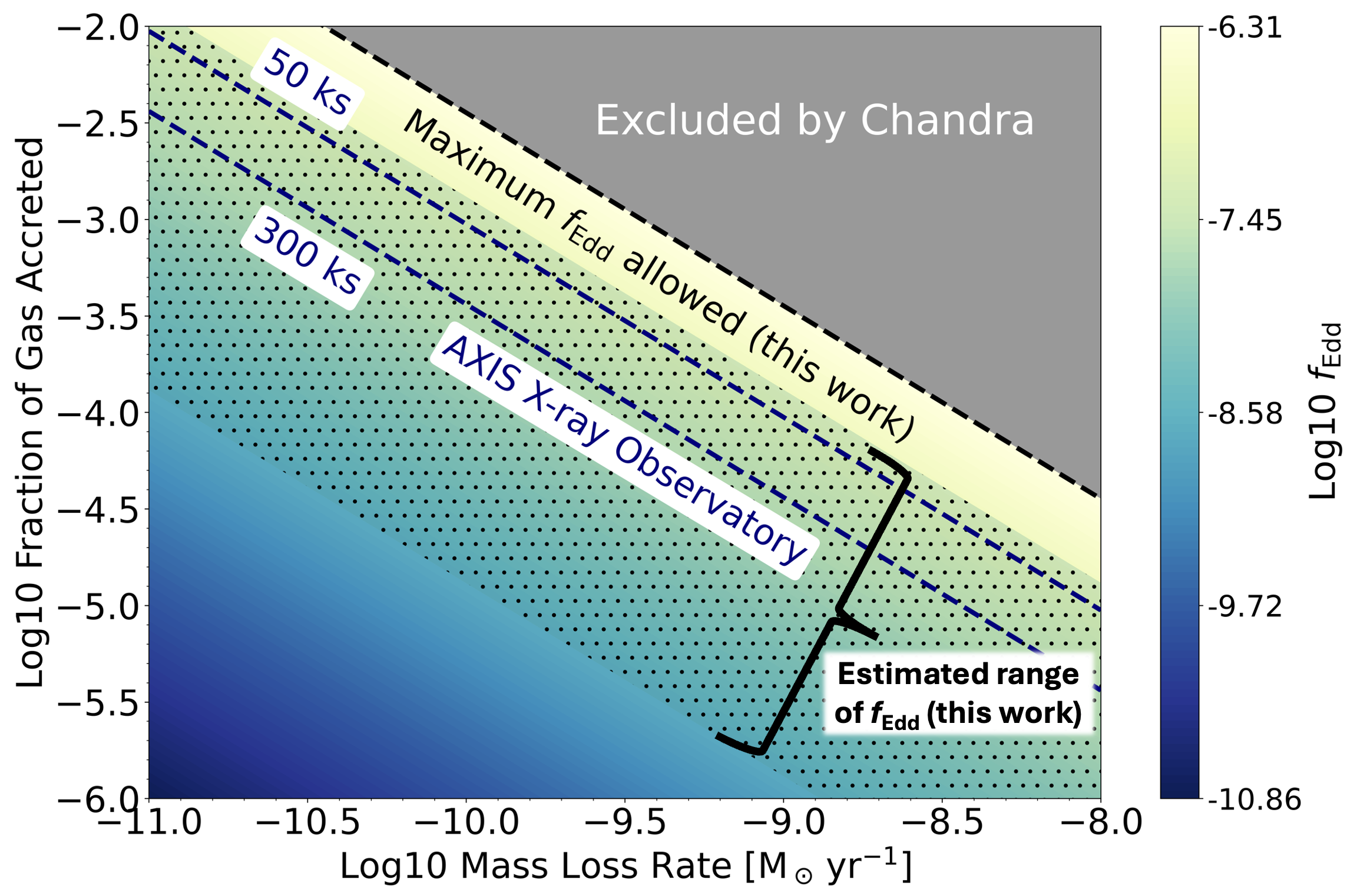}
    \caption{Investigation of the parameter space established by the mass-loss rate and the fraction of gas accreted, which univocally determine the Eddington ratio $\fedd$. ADAF SEDs determine the Eddington ratios our current Chandra $10$ ks observation probed and what a $50$ ks and $300$ ks observation with AXIS \citep{AXIS_2023} could achieve. The hatched region indicates the conservative range of Eddington ratios estimated for Gaia BH3 at its apastron (i.e., its current configuration). Note that at the periastron, we expect Eddington ratios $\sim 50$ times higher, placing most of the parameter space within easy reach of AXIS.}
    \label{fig:contour}
\end{figure}

In Fig. \ref{fig:contour}, we show an analysis of the parameter space ${\cal \dot{M}}$ and ${\cal F}$, which encompasses conservative ranges of these parameters. The computed Eddington ratio $\fedd$ is then translated to an X-ray flux via the ADAF SED model above \citep{Mahadevan_1997, Pesce_2021}.
In this way, we can visualize both the upper limit derived with our current $10$ ks Chandra observation and hypothetical future observations with, e.g., AXIS \citep{AXIS_2023}. The integration times (i.e., $50$ ks and $300$ ks) for AXIS were chosen to be compatible with the wide-area and intermediate-area surveys planned with AXIS \citep{Cappelluti_AXIS_2023}.
In the same Figure, we also show the conservative range of Eddington ratios obtained in Sec. \ref{subsec:theory_estimates}. 
Hence, a modest (i.e., $50-300$ ks) observation with AXIS could easily detect this source even at the apastron of its orbit.

\begin{figure}[h]
   \centering
\includegraphics[width=0.45\textwidth]{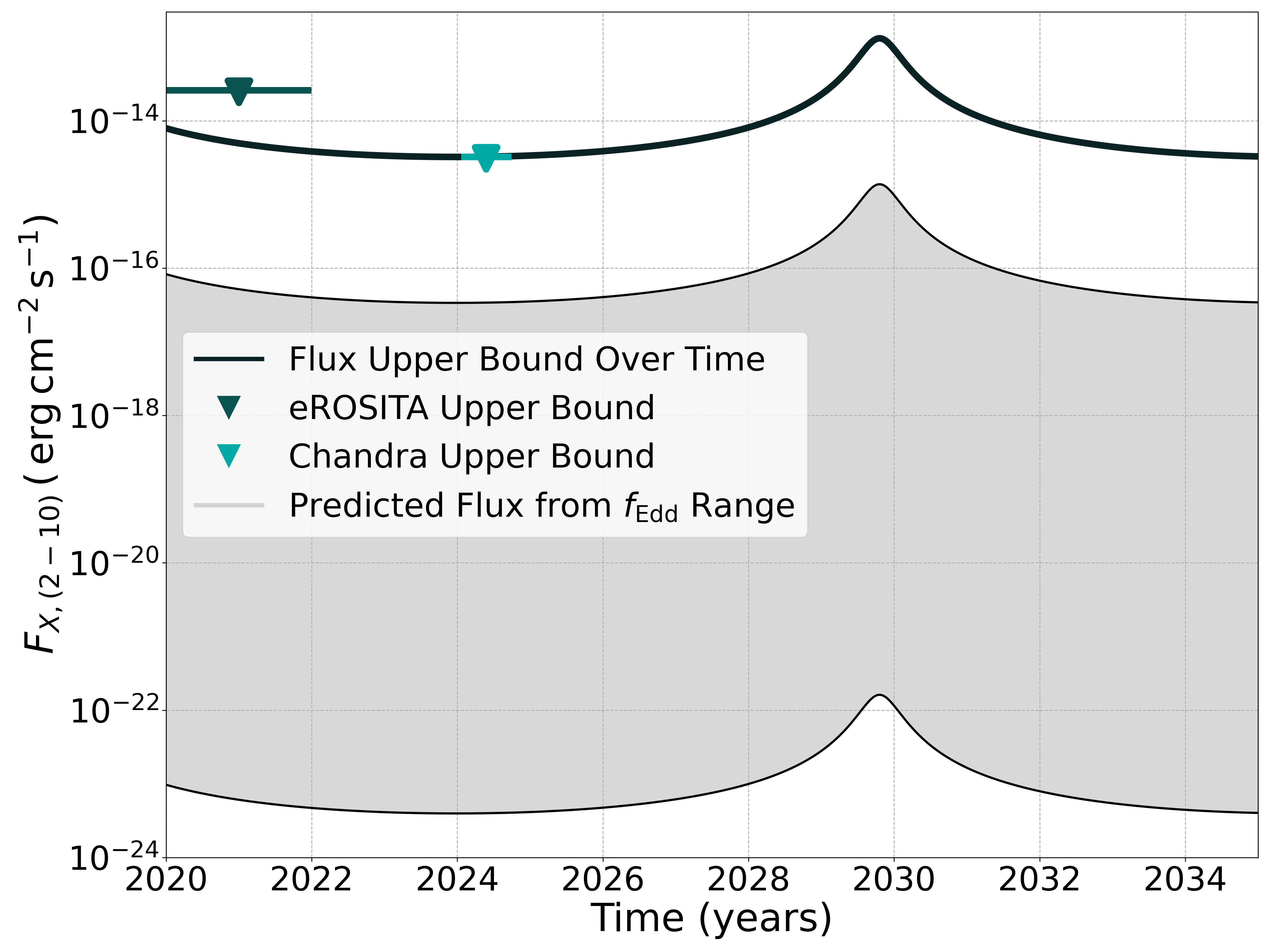}
    \caption{Time variation of the [2-10] keV flux of Gaia BH3 due to the orbital motion of the companion star. The baseline flux is assumed as the upper limit determined by our observation, which is indicated along with the shallower upper limit from eROSITA. The Chandra observation was executed at a time of minimum flux, whereas the peak flux will occur in October 2029. The gray area shows the theoretically predicted range of fluxes, given the range of Eddington ratios estimated in \ref{subsec:theory_estimates}.}
    \label{fig:lc}
\end{figure}

If Gaia BH3 cannot be detected in the X-ray now, it may well be in the future.
Using the best Gaia astrometric solution, the source was observed when the system was close to the apastron, i.e., with an estimated distance between the BH and its companion star of $d_{\star \bullet} \approx 27.86$ AU. Therefore, our observation was performed around the minimum in flux. As the accretion rate scales as $d_{\star \bullet}^{-2}$ (see Eq. \ref{eq:Bondi_new}), we expect the value of $\fedd$ and, consequently, of the flux to increase over time, reaching values $\sim 50$ times higher in $\sim 5$ years.
In Fig. \ref{fig:lc}, we show the evolution of the upper bound of the [2-10] keV flux over time compared with current observational limits. In addition, we display the range of X-ray fluxes estimated, based on the range of $\fedd$ theoretically predicted in \ref{subsec:theory_estimates}. Given our current upper bound, we predict that Gaia BH3 could brighten to up to F$_{2-10}\sim$2$\times$10$^{-15}$ erg cm$^{-2}$ s$^{-1}$ in October 2029 hence detectable with most of the currently operating X-ray telescopes. Long-term transients like this could be detected by an instrument surveying the sky for several years, like eROSITA originally planned, or by combining observations from several observatories' archives. Chandra and XMM-Newton, with the exception of the Galactic Center and deep survey locations, do not offer such an option on a sufficiently large area of sky, making a serendipitous X-ray detection of such a transient very unlikely. However, if a similar system could be found with a few hundred parsecs, instruments like NICER or MAXI could detect them given their likely bright flux.

\section{Summary}

In this paper, we report the most sensitive upper limit on Gaia BH3's X-ray emission and model the parameters of its accretion properties in the Bondi-Hoyle-Lyttleton scenario. 

\begin{itemize}
    \item Gaia BH3 was not detected with a $10$ ks Chandra exposure corresponding to an intrinsic [2-10] keV luminosity of L$_X< 2.10 \times 10^{29} \, \rm erg \, s^{-1}$. 
    \item  For the companion star we estimated mass-loss rates of  10$^{-11} \, \Msun \, \mathrm{yr}^{-1}<$ $\dot{M}_{\text {w}} <  10^{-9} \, \Msun \, \mathrm{yr}^{-1}$ and typical wind velocity $v_{\text {wind}} \sim 200 \mathrm{~km} \mathrm{~s}^{-1}$, leading to a range of Eddington ratios of $ 10^{-9} < \fedd < 10^{-7}$. 
    \item  From the Chandra non-detection, we constrained the Eddington ratio to  $\fedd < 4.91\times 10^{-7}$; therefore, we conclude that any wind-driven accretion over the history of the system contributed to a growth of Gaia BH3 of $\Delta \Mblack \ll 1 \Msun$. This confirms that Gaia BH3 formed near its current mass from a metal-poor star. 
    \item Due to the highly eccentric orbit of the system, we expect that Gaia BH3 will brighten by a factor of $\sim 50$ by 2029, hence greatly enhancing the probability of detection with current and future X-ray facilities.
\end{itemize}

Future observations with Chandra or other facilities may shed more light on the accretion process of this peculiar object. Detecting the electromagnetic emission from Gaia BH3 will be fundamental to informing models for stellar winds and accretion disk models in the lowest Eddington ratio regime thus far explored. Observations with future facilities like the proposed mission AXIS \citep{AXIS_2023} could provide detailed insights on radiatively inefficient accretion flow physics on larger samples of objects.
In addition, observations at longer wavelengths, such as in radio and sub-millimeter with, e.g., VLA and ALMA, could provide additional constraints on the accretion process of this object. Given our model SED, we predict upper limits on the radio and sub-millimeter emission of Gaia BH3 in the range $\sim 1-10 \mu$Jy at apastron (i.e., currently). Recent VLA observations have reported an upper limit consistent with our estimate \citep{VLA_upper_2024}.

Moreover, it is essential to mention that the companion star in Gaia BH3's system is still undergoing significant stellar evolution, currently at the base of the giant branch. As the star expands from its present radius of $\approx 5 R_\odot$ and climbs the giant branch, the accretion rate onto the compact object will likely increase, but on timescales longer than the orbital period. Hence, while future stellar evolution may eventually increase the BH mass, our work proved that past accretion did not significantly contribute to it over the system's lifetime.

The broad importance of such objects, ranging from stellar astrophysics to accretion disk physics, passing through stellar evolution, dark matter, and gravitational waves, warrants more focus on Gaia BH3.

\begin{acknowledgments}
We are grateful to the anonymous referee for helpful suggestions and comments.
N.C., F.P., and G.H. thank the CXC for a prompt turnaround in this DDT observation. 
F.P. acknowledges insightful discussions with Dominic Pesce and Daniel Palumbo regarding ADAF models.
N.C. acknowledges support from the University of Miami College of Arts and Science.
F.P. acknowledges support from a Clay Fellowship administered by the Smithsonian Astrophysical Observatory. This work was also supported by the Black Hole Initiative at Harvard University, which is funded by grants from the John Templeton Foundation and the Gordon and Betty Moore Foundation. 
\end{acknowledgments}

%

\vspace{5mm}
\facilities{Gaia, Chandra (ACIS-S)}


\software{astropy \citep{2013A&A...558A..33A,2018AJ....156..123A}, 
          Source Extractor \citep{Source_extractor}
          }



\bibliography{ms}{}
\bibliographystyle{aasjournal}



\end{document}